# Improved Vehicle Sub-type Classification for Acoustic Traffic Monitoring

Mohd Ashhad
*Dept. of Computer Science & Engineering*
*Jamia Hamdard*
New Delhi, India
ashhad.jamiahamdard@gmail.com

Umang Goenka
*Dept. of Information Technology*
*IIIT Lucknow*
Lucknow, India
lit2019033@iiitl.ac.in

Aaryan Jagetia
*Dept. of Information Technology*
*IIIT Lucknow*
Lucknow, India
lit2019045@iiitl.ac.in

Parwin Akhtari
*Dept. of Mechanical Engineering*
*NIT Rourkeela*
Roorkela, India
parwinakhtari7@gmail.com

Sooraj K. Ambat
*Naval Physical & Oceanographic Laboratory*
*Defence R&D Organization*
Kochi, India
sooraj.npol@gov.in

Mary Samuel
*Dept. of Mathematics*
*IIIT Lucknow*
Lucknow, India
marysamuel@iiitl.ac.in

*Abstract*—The detection and classification of vehicles on the road is a crucial task for traffic monitoring. Usually, Computer Vision (CV) algorithms dominate the task of vehicle classification on the road, but CV methodologies might suffer in poor lighting conditions and require greater amounts of computational power. Additionally, there is a privacy concern with installing cameras in sensitive and secure areas. In contrast, acoustic traffic monitoring is cost-effective, and can provide greater accuracy, particularly in low lighting conditions and in places where cameras cannot be installed. In this paper, we consider the task of acoustic vehicle sub-type classification, where we classify acoustic signals into 4 classes: car, truck, bike, and no vehicle. We experimented with Mel spectrograms, MFCC and GFCC as features and performed data pre-processing to train a simple, well optimized CNN that performs well at the task. When used with MFCC as features and careful data pre-processing, our proposed methodology improves upon the established state-of-the-art baseline on the IDMT Traffic dataset with an accuracy of 98.95%.

*Index Terms*—Acoustic Traffic Monitoring (ATM), Signal Processing, Audio Classification, Machine Learning, CNN

## I. INTRODUCTION

Due to the increasing rates of urbanisation and the associated challenges, traffic monitoring has emerged as an important research area in recent years. It can be used for various applications in smart cities, such as controlling traffic signals, detecting traffic density, predicting and detecting road accidents, etc. Acoustic traffic monitoring (ATM) techniques may help in addressing certain issues that exist in other traffic monitoring methods. Apart from being cost-effective, ATM methods also have the added benefit of preserving the privacy of drivers and vehicles, which might be of vital importance, particularly in sensitive cantonment locations as opposed to vision-based approaches.

ATM algorithms come with their own set of challenges. In the case of acoustic vehicle sub-type classification, it is exceedingly difficult to describe a vehicle using just audio because the sound is a blend of different parts, including the engine, tyres, wind turbulence, outside elements like the road condition. This can occasionally work against the model since it may encounter a variety of complex situations where one of the sound components predominates over another. For instance, there are significant differences between the sounds of a car recorded at a distance of 3 or 5 metres away from the microphone. Additionally, there might also be differences between the sounds a car makes on a dry or wet road.

In this paper we consider the task of acoustic vehicle sub-type classification. Specifically, we investigate a four-class vehicle classification scenario that comprises three vehicle types—cars, trucks, and motorcycles—as well as a no-vehicle class (which contains background noise). Our main contributions through the paper are as follows:

- We show that simple, well optimised CNNs along with simple feature extraction methods can outperform complex, state of the art neural networks such as VGGnets for the task of acoustic vehicle classification.
- We studied the reason behind the poor classification accuracy between the car and truck instances of an open benchmark dataset (IDMT Traffic) [1] and conducted a human performance evaluation to further identify the problem.
- Finally, our proposed pipeline performs better at the task of vehicle sub-type classification with an accuracy of 98.95% with an increase of 42% in the F1 score of the truck class as compared to the established baseline [1].

The ATM models and algorithms that have been the subject of current and pertinent research are briefly described in Section II. The Dataset used for the study has been described in Section III. Section IV explains various approaches employed in the paper. Section V discusses the numerous experiments carried out during the study and provides a thorough analysis of the findings. The work is concluded in Section VI, which also offers suggestions for future research.

## II. RELATED WORKS

Vehicle detection is one of the most fundamental tasks in traffic monitoring. Kazuo et al. [2] developed an ultra low powered acoustic vehicle detection system that used two microphones. Their detector achieved a precision and recall of 0.94 and 0.95 respectively. Shigemi Ishida et al. [3] developed a steady-noise suppression method for better SNR ratio in vehicle detection. Their method achieved F-measures of 0.92 and 0.90 in normal and heavy rain respectively. Billy Dawton et al [4] showed that stereo microphone-based systems perform better as compared to single-microphone systems for vehicle detection and classification. This is why most of the time stereo microphones [1] [6] or multi-channel audio [1] [7] [5] recording is preferred. G.Szwoch et al. [5] used sensors in two directions: parallel and perpendicular to the road and achieved an F1 score of 0.95 for vehicle detection using multiple features.

As for vehicle type and sub-type classification, Amir Y. Nooralahiyan et al. [6] used a Time Delay Neural Network (TDNN) for classifying motorcycles; and light goods vehicles or vans for audios recorded from urban roads in the city of Leeds. They achieved a remarkable testing accuracy of 82.4%. Jobin George et al. [11] considered a 4 way classification scenario: heavy, medium, light, and horns. They used smoothed log energy to detect vehicles and then extracted MFCCs from fixed regions around detected peaks to train an ANN and achieved an accuracy of 67%. An-Chih Yang et al. [7] used a smartphone to record vehicle acceleration at an isolated stop sign in order to avoid diverse traffic and oncoming vehicles, resulting in a dataset with clear audios. Their proposed pipeline comprises both spectral and temporal feature extraction approaches, as well as techniques like noise injection and pitch change to augment the data, which is then sent into several CNN models for classification of vehicles into the following classes: Hybrid, sedan, pickup, bus and commercial .They achieved an accuracy of 75%. Billy Dawton et al. [4] achieved an accuracy of 95.01% on a 3 way classification between : scooter, car and busses using an SVM with a linear kernel. Jakob Abeber et al. [1] provided a publicly available dataset called IDMT Traffic for ATM. They used mel-spectrograms for vehicle sub-type classification and used state of the art CNNs such as VGGnet and Resnets for classification. However their classifier struggled to classify trucks with an F1-score of 0.5.

There are numerous methods that employ computer vision methods as opposed to acoustic features. In the case of images, the model may get visual elements such as headlights, vehicle design, and measurements. There have been several studies that utilised photos or frames in a video to classify vehicle types [8] [9]. Furthermore, several modern techniques, such as infrared thermography [10], may be used to categorise vehicle types based on the thermal properties of vehicle exteriors. Images produce good results, but they are often expensive and difficult to install. Furthermore, there is a privacy concern involved in sensitive areas.

## III. DATASET

For this study, we utilize the IDMT- Traffic dataset [1]. Released in 2021, IDMT Traffic is an open benchmark dataset for acoustic traffic monitoring. The dataset contains time-synchronized stereo audio recordings of moving automobiles made at four separate recording locations, including three city traffic locations and one rural road location in and around Ilmenau, Germany, using both high-quality sE8 microphones and less expensive microelectromechanical systems (MEMS) microphones. Different speed restrictions (30, 50, and 70 km/h), as well as dry and wet road conditions, are included in the recording scenarios. The dataset contains the following: Cars (3903 events), Trucks (511 events), Busses (53 events), and Motorcycles (251 events) as well as background recordings when no vehicles are present.

## IV. METHODOLOGY

### A. Feature extraction and audio pre-processing

For our study, we utilize Mel-Spectrograms, Mel Frequency Cepstral Coefficients (MFCC) and Gammatone frequency cepstral coefficient (GFCC) as main feature extraction methods from the raw audio wave forms that are later on passed on to our CNN based classifier. These methods are usually designed to mimic human audio perception. The stereo audio files' left and right channels were initially processed by averaging them into a mono channel at the unaltered sampling rate of 48 kHz which were later down sampled to the rate of 22.05 kHz during further feature extraction. The finer implementation details have been presented in the following sections.

*1) Mel Spectrograms:* For our first feature extraction method, we computed Mel-Spectrograms using the Librosa python library [12]. A mel-spectrogram computes its output by dividing frequency-domain values by a filter bank and logarithmically depicts frequencies over a predetermined threshold (the corner frequency). We used FFT size of 2048, window size of 1024 and hop size of 512. We kept the number of mel-bands as 128 and applied log magnitude scaling. Fig. 1 depicts Mel-powered spectrograms of samples belonging to different classes from our dataset.

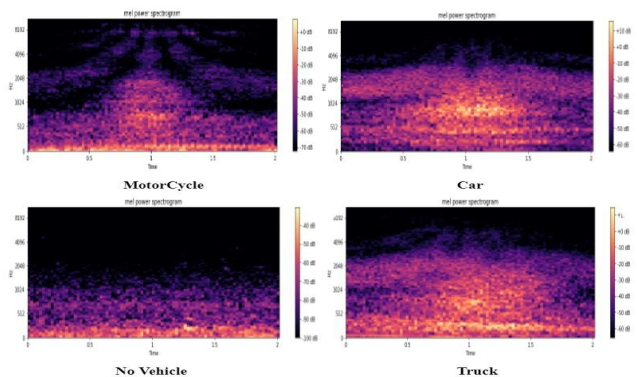

Fig. 1: Mel-Spectrogram of Motorcycle, Car, No Vehicle and Truck

*2) MFCC:* An MFC is made up of a number of coefficients known as mel-frequency cepstral coefficients (MFCCs). They were created using an audio clip's cepstral representation (a nonlinear "spectrum-of-a-spectrum"). The mel-frequency cepstrum (MFC) differs from the cepstrum in the fact that the frequency bands are evenly spaced on the mel scale, which more closely resembles the response of the human auditory system than the linearly-spaced frequency bands used in the conventional spectrum. To compute the MFCCs for the samples in our dataset, we used the Librosa Python Library [12]. We used the number of MFCCs to return as 128. The size of the FFT, window length and hop length were kept to be the same as in the previous section.

*3) GFCC:* A more thorough model was created to imitate the properties of the ear based on psychophysical investigations of the auditory peripheral after the widespread adoption of Mel Frequency Filters because of its similarities to the human ear model. Patterson and Smith [15] created the GFCC (gammatone frequency cepstral coefficient). The human auditory systems are represented by the Gammatone filter bank as a set of overlapping band-pass filters. Equation (1) gives each filter's impulse response.

$$g(t) = at^{n-1}e^{-2\pi bt}cos(2\pi f_c t + \phi) \quad (1)$$

where $f_c$ is the central frequency, $a$ is a constant (often equal to 1), $n$ is the order of the filter, $\phi$ is the phase shift, and $b$ is the bandwidth. The filter's Equivalent Rectangular Bandwidth (ERB) can be used to determine the centre frequency and bandwidth. We used the Spafe Python library [13] to extract GFCC features. We extracted 64 cepstrums with the number of filters set to be 128.

### B. Data augmentation

We used temporal stretching as our primary data augmentation strategy to obtain more training/testing samples to reduce data imbalance issues from some of our experiment variations (see Section V).

*1) Temporal stretching and zero padding:* Time stretching is a simple augmentation method where the voice pitch is unaffected while the recording period is extended (the glottal tone frequency is not changed). A 100% increase in playback time is possible (two times the original playback time). If the stretch factor is greater than 1 then the signal is sped up and if the factor is less than 1 then the signal is slowed down. We used the Librosa library to implement time stretching. We used a stretching factor of 1.5, 0.9 and 1.2 depending on the variant of our experiment (refer Section V). After applying time stretching on a signal, we applied zero padding (post padding) so as to keep the length of each recording constant at 2 seconds to avoid any dimensional issues that that the model might run into.

### C. Classification model

After careful tuning and evaluation, we came up with a relatively simple Convolutional Neural Network based classification model which performs well at the task. Our model consists of four 1D Convolutional layers each of which are followed by a 1D Maxpool layer. The output from the final Maxpool layer is then passed on to a flatten layer which further feeds the output of the previous layer into the following three dense layers that allows the final softmax layer to provide the required prediction. We also used dropout regularization to allow our model to generalize better (dropout probability = 0.3). We trained our model for 30 epochs along with learning rate reduction (minimum lr = 1.0e-05) with a patience of 2 and Adam optimizer. The computational diagram of our classification model is depicted in Fig. 2.

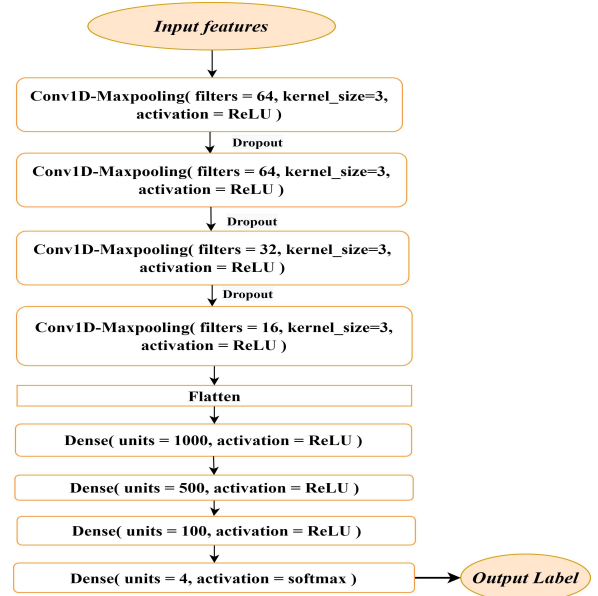

Fig. 2: Layer Diagram of our CNN Model

## V. EXPERIMENTS AND ANALYSIS

All the experiments were performed on Google Colab Pro with 32Gb of main memory, Nvidia A100 GPU with 16Gb of GPU Ram and Intel Xeon CPUs (subjective to availability).

### A. Reproducing baseline results

Since the code was not made available in the IDMT Traffic study [1], we attempted to replicate their methods and baseline results as closely as possible. We kept the number of mel bands at 16 and extracted features using mel-spectrograms. According to the train-test split (refer Table II) of authors in [1], samples of vehicles travelling at speeds of 30 km/h and 50 km/h were utilised for training, while those of travelling at 70 km/h were used for testing. We obtained fairly comparable outcomes in three of the four classes (refer Table I), however we were unable to obtain an F1-score of 0.5 for the truck class, as the authors claimed.

### B. Establishing results using a simpler model and varied feature extraction methods

We established vehicle sub-type classification results on the dataset using our CNN-based classification model. Addition-ally, we experimented with various feature extraction techniques to determine which one was best for our application. Our tests show that Mel-Spectrograms performs the best, with an accuracy of 94.24% (refer Table III). While employing less than half the amount of trainable parameters, our simpler CNN model produced outcomes comparable to those of the VGG model used in [1] and Section V-A. However, the classifier still performed poorly in the truck class with an F1 score of 0.35.

TABLE I: Class-wise F1 score of the baseline and our replication of the results

| Model | Car | Truck | Motorcycle | No Vehicle |
|---|---|---|---|---|
| Abeßer, J et al. [1] | 0.94 | 0.5 | 0.96 | 1.00 |
| Our replication | 0.94 | 0.35 | 0.95 | 1.00 |

TABLE II: Train/validation/test split used by authors in [1]

| Class | Training Samples | Validation Samples | Testing Samples |
|---|---|---|---|
| Car | 2471 | 275 | 1157 |
| Motorcycle | 132 | 15 | 99 |
| No Vehicle | 2393 | 266 | 1412 |
| Truck | 290 | 32 | 189 |

TABLE III: Classification results of our model with varied feature extraction methods

| Class | Feature | Precision | Recall | F1-Score | Accuracy |
|---|---|---|---|---|---|
| Car | MFCC | 0.87 | 0.96 | 0.91 | 93.41% |
| Motorcycle | | 0.97 | 0.89 | 0.92 | |
| No Vehicle | | 0.99 | 0.98 | 0.98 | |
| Truck | | 0.65 | 0.21 | 0.31 | |
| Car | GFCC | 0.88 | 0.96 | 0.92 | 93.78% |
| Motorcycle | | 0.96 | 0.91 | 0.93 | |
| No Vehicle | | 0.99 | 0.99 | 0.99 | |
| Truck | | 0.65 | 0.23 | 0.33 | |
| Car | Mel-Spectrogram | 0.88 | 0.98 | 0.93 | **94.24%** |
| Motorcycle | | 0.98 | 0.90 | 0.94 | |
| No Vehicle | | 1.00 | 0.99 | 1.00 | |
| Truck | | 0.69 | 0.22 | 0.35 | |

### C. Classification with data balancing and data augmentation

The classifier performed poorly in the truck class irrespective of the feature extraction methodology utilized (refer Table III) highlighting a possible data imbalance issue. Consequentially, we experimented with a balanced training technique to get rid of the dataset's class imbalance issue. Additionally, we used temporal stretching as a method of data augmentation to double the number of training samples for the motorcycle and truck classes at speeds of 30 km/h and 50 km/h, from 147 and 322 to 294 and 644, respectively. 700 random samples were selected from the car and no vehicle classes. The test samples used were the same as in section V-A and V-B. By using this technique, we were able to marginally boost the truck class F1-score from 0.35 to 0.41 without significantly influencing the performance of the motorcycle or no vehicle class, however there was a clear decline in F1-score of the car class (refer Table IV).

TABLE IV: Classification accuracy with data balancing and augmenting

| Class | Precision | Recall | F1-Score |
|---|---|---|---|
| Car | 0.96 | 0.64 | 0.77 |
| Motorcycle | 0.89 | 0.84 | 0.86 |
| No Vehicle | 0.99 | 0.99 | 0.99 |
| Truck | 0.27 | 0.84 | 0.41 |

### D. Human performance evaluation

In order to further understand the IDMT Traffic dataset and the associated problems, we performed a human performance study where 20 different subjects (all between ages of 20-25) were made to listen to 20 audios each (5 samples from each class) from the dataset. These audios were randomly sampled from the 70 km/h vehicle recordings so as to resemble the testing set of the Neural Network used in the previous subsections. The subjects were asked to label each of these unlabeled audios in one of the four classes without any previous training. The results of this study have been summarized in Table V. The results show that over 60% of the truck samples were misclassified as cars suggesting a possible problem with the dataset, particularly in the truck class.

TABLE V: Confusion matrix of human study (values in %)

| | Car | Truck | Motorcycle | No Vehicle |
|---|---|---|---|---|
| Car | 95 | 3 | 1 | 1 |
| Truck | 61 | 38 | 1 | 0 |
| Motorcycle | 3 | 1 | 95 | 1 |
| No Vehicle | 1 | 0 | 1 | 98 |

### E. Study of misclassified truck samples

In order to evaluate the reason behind the high level of misclassifications of trucks as cars, we analyzed the wrongly classified samples of trucks by both the Neural Network as well as the human subjects. It was discovered that there was a correlation between the misclassifications such that over 90% of the wrongly classified truck samples by the human subjects were misclassified by the Neural Network as well. Signal analysis of the correctly classified and misclassified samples of the truck class revealed a common trend. The Root Mean Square Energy of the correctly classified samples was much higher as compared to their misclassified counterparts (refer Fig. 5). Similarly, In the misclassified samples, the magnitude of all the pure tone frequencies as calculated by the Fast Fourier Transform was very low (refer Fig. 3 and Fig. 4) as compared to the correctly classified truck samples.

Both these phenomenons could be explained by the distance of the passing trucks from the recording microphone such that, when the distance of the passing truck is less from the microphone, the microphone is better able to record higher spectrum of frequencies at a greater amplitude, hence RMSE is higher along with higher magnitudes of pure tone frequencies as opposed to when the truck is passing at a greater distance from the microphone. We argue that the samples of trucks with low values of RMSE are unreliable and cannot be used for training and evaluation of the model as they don't provide sufficient features for vehicle sub-type classification. The high order of difference between the RMSE values between samples belonging to the same class was only observed in the truck samples (refer Fig. 6 and Fig. 7). To further verify our analysis, we used an unsupervised approach to first separate the two types of truck samples and then perform vehicle sub-type classification. The two types of samples are as follows: 1) A good sample, thus higher magnitudes of RMSE 2) A bad sample, thus lower magnitude of RMSE. We calculated the RMSE values for each of the samples in the truck class using equation (2).

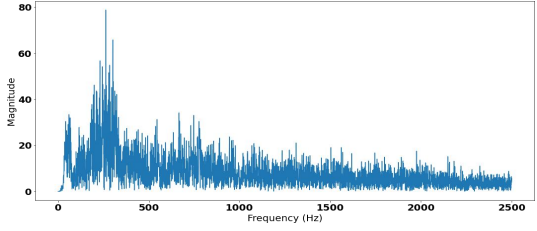

Fig. 3: Spectrum of a correctly classified truck sample

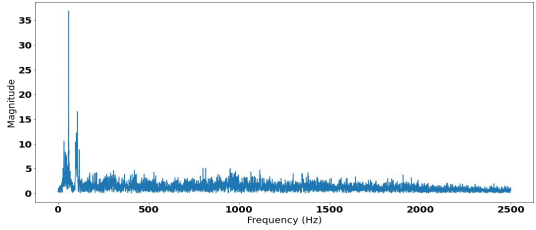

Fig. 4: Spectrum of a misclassified truck sample

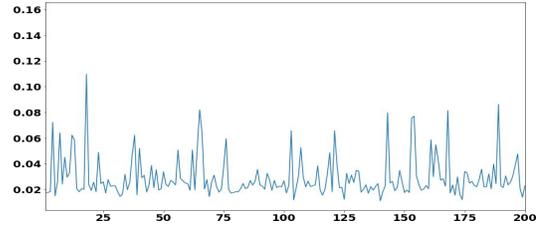

Fig. 6: RSME values of 200 random truck recordings

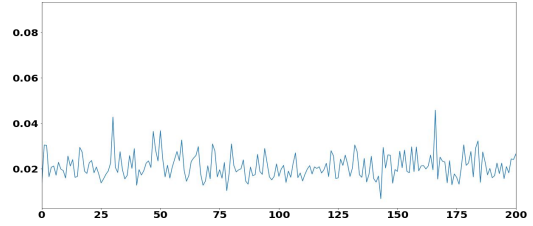

Fig. 7: RSME values of 200 random car recordings

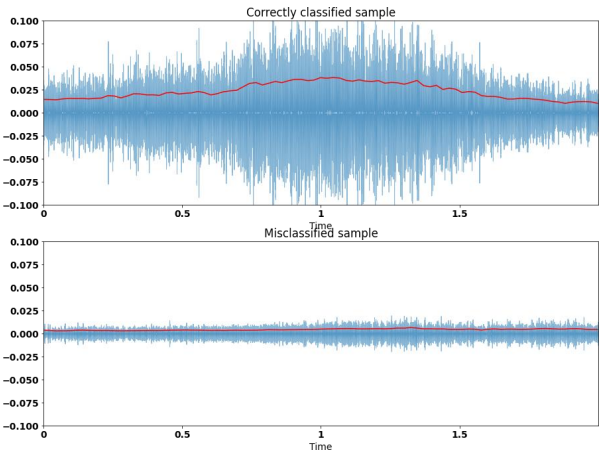

Fig. 5: Frame-wise RSME values of a correctly classified and a misclassified truck sample (shown in red)

$$\sqrt{\frac{1}{K} \sum_{k=t.K}^{(t+1).K-1} s(k)^2} \quad (2)$$

where $K$ is the number of samples in a frame and $t$ is the frame number. We used these values to create a vector of length 87 (All of our recordings were of 2 second each at the sampling rate of 22.05 kHz). These vectors were then passed into a K-Means clustering algorithm with the number of clusters set to 2. The number of samples in cluster 1 and 2 were computed to be 444 and 67 respectively. Signal analysis revealed that cluster 1 contained the majority of samples with low magnitude of RMSE (bad samples) while cluster 2 contained the majority of samples with higher magnitude (good samples).

### F. Vehicle sub-type classification after cleaning the dataset

After removing the bad samples from the truck class we again performed vehicle sub-type classification. Since there weren't enough samples of trucks for effective training and testing, we used time stretching to increase the number of high-quality samples to obtain a total of 268 samples, we applied stretching factors of 1.5, 0.8, and 1.2 to our chosen samples. In order to properly calibrate the model for real-world applications where the speeds of the vehicles can vary, we then picked all the motorbike samples from the dataset together with all the car and no vehicle audio samples as used in section V-A and V-B, independent of the speeds. We divided this data into a 70:30 train-test split after randomly shuffling them. Our model achieved an overall accuracy of 98.95% along with a 42% increase in the F1-score of the truck class when combined with MFCC as a feature extraction method (refer Table VI and VII). The pipeline of our proposed methodology has been depicted in Fig. 8.

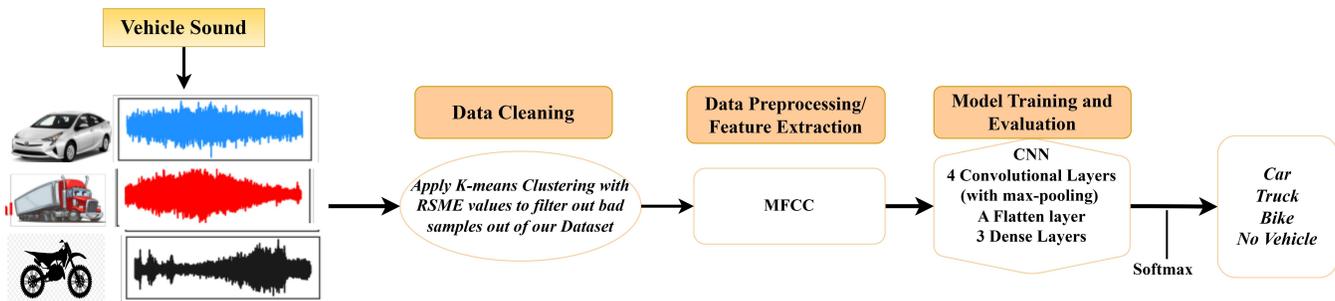

Fig. 8: The pipeline of our proposed methodology. The captured data is first being cleaned where samples are rejected based on their RMSE values. MFCC features are then extracted from the remaining samples and are passed on to the CNN for training and evaluation.

TABLE VI: Classification results after removing bad samples and using varied feature extraction methods

| Class | Feature | Precision | Recall | F1-Score | Accuracy |
|---|---|---|---|---|---|
| Car | MFCC | 0.99 | 0.99 | 0.99 | **98.95%** |
| Motorcycle | | 1.00 | 0.97 | 0.99 | |
| No Vehicle | | 0.99 | 1.00 | 1.00 | |
| Truck | | 0.95 | 0.89 | 0.92 | |
| Car | GFCC | 0.98 | 1.00 | 0.99 | 98.94% |
| Motorcycle | | 0.97 | 0.99 | 0.98 | |
| No Vehicle | | 1.00 | 0.99 | 0.99 | |
| Truck | | 0.97 | 0.84 | 0.90 | |
| Car | Mel-Spectrogram | 0.98 | 0.99 | 0.98 | 98.08% |
| Motorcycle | | 0.89 | 0.96 | 0.93 | |
| No Vehicle | | 1.00 | 0.99 | 0.99 | |
| Truck | | 0.86 | 0.83 | 0.84 | |

TABLE VII: Comparison of class-wise F1 score of our model with [1]

| Model | Car | Truck | Motorcycle | No Vehicle |
|---|---|---|---|---|
| Abeßer, J et al. [1] | 0.94 | 0.5 | 0.96 | 1.00 |
| **Our Model** | **0.99** | **0.92** | **0.99** | **1.00** |

## VI. CONCLUSION AND FUTURE WORKS

Acoustic traffic monitoring is a promising field and further research is needed in it for the ever increasing demand of smart cities. Through our paper and analysis, we were able to show that simple, well optimised CNNs along with careful data pre-processing can often provide either comparable or better results than their more complex and computationally expensive counterparts in the task of acoustic vehicle subtype classification, thus increasing the utility of our approach in real world deployment. Using Mel Frequency Cepstral Coefficients and our simple CNN model as well as sample rejection method, we were able to outperform the established baseline on the IDMT Traffic dataset in all the classes, with a gain of **42%** in the F1-score of the truck class. In section V-F, we hypothesize a possible correlation between the distance of the recording microphone from the vehicle and the quality of features recorded, thus affecting the classification accuracy. In further research, we want to explore this correlation to find the optimal quality of microphone and the recording distance needed for real-world mass deployment.

The Code for the above experiment is made publicly available [15].